\documentclass[12pt,a4paper]{article}
\usepackage{epsfig}

\newcommand{\Redtau}{{\rm Re}\ d_\tau}
\newcommand{\Imdtau}{{\rm Im}\ d_\tau}
\newcommand{\ecm}   {e-cm}
\newcommand{\TauP}  {\ensuremath{\tau \rightarrow \pi\,\nu_\tau}}
\newcommand{\TauRho}{\ensuremath{\tau \rightarrow \rho\,\nu_\tau}}

\newcommand{\tauppp}{\ensuremath{\tau^- \rightarrow \pi^-\,\pi^+\,\pi^-\,\nu_\tau}}
\newcommand{\taupppn}{\ensuremath{\tau^- \rightarrow \pi^-\,\pi^0\,\pi^0\,\nu_\tau}}
\newcommand{\taue}  {\ensuremath{\tau^- \rightarrow \mathrm{e}^-\,\overline{\nu}_{\mathrm{e}}\,\nu_\tau}}
\newcommand{\taumu} {\ensuremath{\tau^- \rightarrow \mu^-\,\overline{\nu}_\mu\,\nu_\tau}}
\newcommand{\pio}   {\ensuremath{\pi^0}}
\newcommand{\eetautau} {\ensuremath{e^+e^- \rightarrow \tau^+\tau^-}}

\begin{document}
\begin{flushright}    
  {\today} \\
  IISc-CTS-02/02
\end{flushright}
\bigskip

\begin{center}
{\large
{\bf CP Violation in the Production of {\boldmath $\tau$}-Leptons \\
     at TESLA with Beam Polarization\raisebox{0cm}[0.6cm]{}}}

\vskip 2cm

{\bf B.\ Ananthanarayan}\\
Centre for Theoretical Studies, Indian Institute of Science, \\
Bangalore 560 012, India\\

\bigskip

{\bf Saurabh D.\ Rindani}\\
Theory Group, Physical Research Laboratory, \\
Navrangpura, Ahmedabad 380 009, India \\

\bigskip

{\bf Achim Stahl}\\
DESY, Platanenallee 6, \\
15738 Zeuthen, Germany

\end{center}

\bigskip

\bigskip

\begin{abstract}
We study the prospects of discovering CP-violation in the production of 
{$\tau$} leptons in the reaction {$e^+ e^- \to \tau^+ \tau^-$} at TESLA, 
an {$e^+e^-$} linear collider with center-of-mass energies of 500 or even 
800 GeV.
Non-vanishing expectation values of certain correlations between the 
momenta of the decay products of the two {$\tau$} leptons would signal
the presence of CP-violation beyond the standard model.
We study how longitudinal beam polarization of the electron and positron 
beams will enhance these correlations.  
We find that T-odd and T-even vector correlations are well suited for
the measurements of the real and imaginary parts of the electric 
dipole form factors.
We expect measurements of the real part with a precision of roughly 
{$10^{-20}$}~{\ecm} and of the imaginary part of {$10^{-17}$}~{\ecm}.
This compares well with the size of the expected effects in many extensions
of the standard model.

\end{abstract}

\newpage

\section{Introduction}
One possible signal for physics beyond the standard model (SM) would
be the presence of significant CP violation in the production
of {$\tau$} leptons~\cite{bbno,bno}, or in their decay \cite{tsai}. 
CP violation in production would arise,
assuming that {$e^+e^-$} annihilate into a virtual {$\gamma$} or $Z$, from the
electric dipole form factor (EDM)
{$d^\gamma_\tau(q^2)$}, or its generalization for the {$Z$} coupling,
the so-called ``weak" dipole moment form factor (WDM) {$d^Z_\tau(q^2)$}. 
These are expected to be unobservably small in the SM. 
Thus, observation of CP violation in production would be unambiguous 
evidence for physics beyond the SM.
The signal for CP violation in $\tau$ production would be the
non-zero values of certain
momentum correlations of the $\tau$ decay products~\cite{bbno,bno}
since the momenta play the role of spin analyzers for the $\tau$ leptons.

TESLA is a proposed $e^+e^-$ linear collider with a 
center-of-mass energy of 
$\sqrt{s}=500$~GeV with the possibility of an extension to
$800$ GeV~\cite{tdr}.  It is a multi-purpose machine that will
test various aspects of the standard model (SM) and search for signals
of interactions beyond the standard model.  
A strong longitudinal polarization program
at TESLA with considerable polarization of the electron beam, with
the possibility of some although not as high a degree of
polarization of the positron beam is also planned \cite{Gudi}.  
We assume an integrated luminosity of a few hundred fb$^{-1}$, 
which would lead to copious production of $\tau^+\tau^-$ pairs.
Longitudinal beam polarization of
the electron and positron beams would lead to a substantial enhancement
of certain vector correlations, which are non-vanishing in
the event that the $\tau$ lepton has an EDM or a WDM.
We will assume integrated luminosities of 500 and 1000 fb$^{-1}$ at center of 
mass energies of 500 and 800 GeV, respectively.
We expect a magnitude of polarization of {$80\:\%$} for the electron beam and 
{$60\:\%$} for the positron beam.

Our aim in this work is to study these vector correlations constructed
from the momenta of the charged decay products 
of {$\tau$} leptons.  We will confine our attention to two-body decays of
the {$\tau$} leptons for which the correlations as well as their variance due 
to CP conserving standard model interactions can be computed analytically. 
We also study the effects of helicity-flip bremsstrahlung~\cite{basdr3}
which contributes to these correlations at $\cal{O}(\alpha)$. 
This is a standard model background to the signal of interest.

The plan of the paper is the following:  in the next section we present 
estimates for the EDM and WDM in certain popular extensions of the SM 
which sets the scale for our studies. 
In Sec.\ \ref{sec:VecCorr}, we discuss the correlations in general and 
reproduce results from the literature for our vector correlations.  
In Sec.\ \ref{sec:TESLARes}, we discuss the numerical results for TESLA 
energies, luminosities, and polarizations, including the limits achievable.
In Sec.\ \ref{sec:Conclusions} we discuss the implications of our results for 
the configuration being planned at TESLA and how our results
will translate into certain design criteria for the machine and the detector.

\section{Dipole Moments in Extensions to the\\ Standard Model}
\label{sec:Extensions}

CP violating dipole moments of leptons can arise in the standard model
radiatively. However, since there is no CP violation in the lepton sector in
SM, it can only be induced by CP violation in the quark sector. One has to go
at least to 3-loop order to generate a non-vanishing contribution to the lepton
dipole moments \cite{bern1,hoogeveen}. A crude estimate gives
\begin{equation}
\vert d_{\tau}(SM)\vert \stackrel{<}{\sim} 10^{-34}\;\mbox{\ecm}.
\end{equation}

Extensions of the SM where complex couplings appear can easily generate
CP-violating dipole moments for {$\tau$} lepton at one-loop order. 
Provided these couplings are generation or mass dependent, 
it is possible that reasonably large dipole moments for the {$\tau$} lepton
are generated, while continuing to satisfy the constraints
coming from strong limits on the electric dipole moments of the electron
or the neutron.

Since dipole couplings of fermions are chirality flipping, they would be
proportional to a fermionic mass. However, this need not necessarily be the
mass of the {$\tau$} lepton. It could be the mass of some other heavy fermion 
in the theory. 
As a result, the dipole coupling at energy of $\sqrt{s}$ need not
necessarily by suppressed by a factor of $m_{\tau}/\sqrt{s}$, but could involve
a factor $m_F/\sqrt{s}$, where $F$ is a new heavy particle in the extension of
the SM, and this factor need not be small. It is thus possible to get 
dipole form
factors almost of the order of $(\alpha/\pi)$ in units of an inverse mass which
appears in the loop. If the mass is that of $W$ or $Z$, it is possible to get
dipole moments of the order of $10^{-19}$ \ecm.
In actual practice, however, the particle appearing in the loop is also 
constrained to be heavy. As a result, the dipole moments in left-right
symmetric models, Higgs exchange model with spontaneous CP violation and
natural flavor conservation, and supersymmetric models turn out to
be of the order of $10^{-23}$ \ecm, or smaller \cite{bern1}.

In the case of most models, information exists in the literature for the values
of electric and weak dipole form factors only at $q^2$ values of 0 and
$m_\tau^2$, respectively. Models in which the $q^2$ dependence of the
CP-violating form factors has been studied are scalar leptoquark models with
couplings only to the third generation of quarks and leptons. 
Of these, the most promising model is the one in which the
leptoquark transforms as an $SU(2)$ doublet. In \cite{bern2} a value of 
Re~$d_\tau^\gamma \stackrel{<}{\sim} 3\cdot 10^{-19}$~{\ecm} was found
above the $Z$ resonance, with Re $d_\tau^Z$ about $\frac{1}{4}$ of this value 
for a more favorable case.
Taking into account restrictions on the doublet leptoquark mass and couplings 
coming from LEP data values of 
$d_\tau^\gamma \stackrel{<}{\sim} 10^{-19}$~{\ecm}
and $d_\tau^Z \stackrel{<}{\sim}(\rm few)\cdot 10^{-20}$~\ecm were estimated
at $\sqrt{q^2} = 500$~GeV \cite{pou}. 

Values of $\tau$ dipole form factors of the order of {$10^{-19}\;\mbox{\ecm}$} 
are also obtained in
models with Majorana neutrinos of mass of a few hundred GeV, and of the order
of $10^{-20}\;\mbox{\ecm}$ in two-Higgs doublet extensions with natural flavor
conservation, and in supersymmetric models through a complex
$\tau-\tilde{\tau}-$neutralino coupling, not far from the $\tilde{\tau}$
threshold \cite{bern2}.

\section{The Vector Correlations}
\label{sec:VecCorr}

In ref.~\cite{bbno} an
extensive analysis of momentum correlations was first presented in
the context of the reaction $e^+ e^- \rightarrow Z^0 \rightarrow
\tau^+ \tau^-$, and was subsequently generalized to energies far
away from the $Z$ resonance in~\cite{bno}.  Expressions were presented
there for the $\tau$ production matrix {$\chi$} involving the
EDM and the WDM, in addition to the main contribution from
SM vertices, which is required to compute the production cross-section,
the momentum correlations of interest, as well as their variances,
also taking into account the ${\cal D}$ matrices that account for
the decay of the $\tau$ into (two-body) final states.  
In ref.~\cite{basdr1} it was shown that in the limit of vanishing
electron mass, the initial state depends on the electron polarization 
{$P_e$} and the positron polarization {$P_{\overline{e}}$} only through
the CP-even combinations $(P_e-P_{\overline{e}})$ and 
$(1-P_e P_{\overline{e}})$.
Therefore one may search for CP violation through the non-vanishing 
expectation values of CP-odd momentum correlations for arbitrary 
electron and positron polarizations.  
In particular, simple ``vector'' correlations
for which the correlations as well as their variances could be computed
in closed form, were shown to have enhanced sensitivity to CP violating
EDM and WDM of the $\tau$ leptons at SLC, and at the energies of a 
Tau-charm factory.  In this work, we shall examine
these correlations for their sensitivity to $\tau$ EDM and WDM
at TESLA energies, polarizations and luminosities.

The CP-odd momentum correlations we consider here are associated 
with the center of mass momenta
${\bf q}_{\overline{B}}$ of $\overline{B}$ and
${\bf q}_A$ of $A$, where the $\overline{B}$ and
$A$ arise in the decays $\tau^+ \rightarrow \overline{B}
\overline{\nu}_{\tau}$
and $\tau^-\rightarrow A \nu_{\tau}$, and where $A,\ B$ run over
$\pi$, $\rho$, $a_1$, etc.  In the case when $A$ and $B$ are
different, one has to consider also the decays with $A$ and $B$
interchanged, so as to construct correlations which are
explicitly CP-odd.  

The correlations we consider are
\begin{equation}\nonumber
 O_1\equiv \frac{1}{2} \left[\hat{\bf p}\cdot \left({\bf q}_{\overline B}\times
{\bf q}_A \right)
+\hat{\bf p}\cdot
\left({\bf q}_{\overline A}\times  {\bf q}_B \right)\right]   
\end{equation}
and 
\begin{equation} \nonumber
O_2 \equiv
\frac{1}{2}\left[\hat{\bf p}\cdot \left({\bf q}_A + {\bf
q}_{\overline B} \right)+\hat{\bf p}\cdot \left({\bf q}_{\overline A} +
{\bf q}_B \right)\right],
\end{equation}
where $\hat{\bf p}$ is the unit vector
in the positron beam direction. 

Note that since $O_2$ is
CPT-odd it measures ${\rm Im}\,d_{\tau}^i$, whereas $O_1$ is CP-even and
measures ${\rm Re}\,d_{\tau}^i$.
An additional advantage of these correlations is
that the correlations as well as the standard deviation for the operators
due to the standard model background are both calculable in closed form
for two-body decays of the $\tau$ leptons.

The calculations include two-body decay modes of the $\tau$ in general
and are applied specifically to the case of {\TauP} and {\TauRho} due to the
fact that these modes possess a good resolving power
of the $\tau$ polarization, parameterized in terms of
the constant $\alpha_\pi=1$ for the
$\pi$ channel (with branching fraction of about
$11\%$) and $\alpha_\rho=0.46$ for the $\rho$ channel (with branching
fraction of about $25\%$) from the momentum of the {$\pi$} or the {$\rho$}.

Analytic expressions for these correlations can be found in \cite{basdr2}. 
For the sake of convenience of reference and completeness, we give the
expressions here.  With the definitions $
r_{ij}\equiv (V_e^i A_e^j + V_e^J A_e^i)/(V_e^i V_e^j + A^e_i A^e_j)$ and the
effective polarization parameter 
$P\equiv (P_e - P_{\overline{e}})/(1- P_e P_{\overline{e}})$
(the vector and axial-vector couplings $A_l^i,\, V_l^i, \,
l=e~\mbox{or}~\tau, \, i=\gamma~\mbox{or}~Z$ are listed in \cite{bno}),
\begin{eqnarray} \nonumber
& \langle O_1\rangle   =  -
{{1}\over{36x\sigma}}
\sum_{i,j}K_{ij}s^{3/2}m_{\tau}^2
(1-x^2)\left(\frac{r_{ij} - P}{1-r_{ij}P}\right) &  \nonumber \\
 & [(A_\tau^i\,\Redtau^j+A_\tau^j\,\Redtau^i)\,
\alpha_A \alpha_B\,(1-p_A)(1-p_B) - & \nonumber \\
& \frac{3}{2}(V_\tau^i\,\Redtau^j+V_\tau^j\,\Redtau^i)\,
[\alpha_A(1-p_A)(1+p_B)+
\alpha_B(1-p_B)(1+p_A)], &
\end{eqnarray}
and
\begin{eqnarray}
& \langle O_2\rangle = {{1}\over{3\sigma}}
\sum_{i,j}K_{ij}\,s^{3/2}\,m_{\tau}
\left(\frac{r_{ij} - P}{1-r_{ij}P}\right) & \nonumber \\
& \frac{1}{4}(A_\tau^{i}\,\Imdtau^j+A_\tau^j\,\Imdtau^{i})(1-x^2)
(\alpha_A(1-p_A)
+\alpha_B(1-p_B)),&
\end{eqnarray}
where $x =
2m_{\tau}/\sqrt{s}$, $p_{A,B}=m_{A,B}^2/m_\tau^2$, and $\sigma$ is
the cross-section of {\eetautau} given by:
\begin{equation}
\sigma  =\sum_{i,j}K_{ij}\,s\,
[V_\tau^iV_\tau^j(1+\frac{x^2}{2})+
A_\tau^iA_\tau^j(1-x^2)],
\end{equation}
and
\begin{equation}
K_{ij}=\frac{ (V_e^iV_e^j+A_e^iA_e^j)(1-r_{ij}P)}
{12 \pi  (s-M_i^2)(s-M_j^2)}(1-x^2)^{1/2}[1-P_e P_{\bar{e}}].
\end{equation}
Since the energies involved are far above the $Z$ resonance, we have neglected
the $Z$ width in the expressions above.

We have analytic expressions for the variance
$S_a^2\equiv \langle  O^2_a\rangle  - \langle  O_a\rangle ^2 
\approx \langle  O^2_a\rangle $
in each case, arising from the CP-invariant SM part of the
interaction:
\begin{eqnarray}\nonumber
  &   \langle O_1^2\rangle  =
{{1}\over{720x^2\sigma  }}\sum_{i,j}K_{ij}s m_\tau^4
 \biggl(
(1-p_A)^2(1-p_B)^2 & \nonumber \\
&
  [V_\tau^iV_\tau^j(6+8x^2+x^4)+
A_\tau^iA_\tau^j(6-2x^2-4x^4)] &  \nonumber \\
  & +(1-x^2)
\left([(1+p_A)^2(1-p_B)^2+(1+p_B)^2(1-p_A)^2] \right. & \nonumber \\
& \left. [3V_\tau^iV_\tau^j(3+2x^2)+9A_\tau^iA_\tau^j(1-x^2)
] \right. & \nonumber \\
& \left. +4\alpha_A\alpha_B
(1-p_B^2)(1-p_A^2)(1-x^2)[V_\tau^iV_\tau^j
-A_\tau^iA_\tau^j]\right)   & \nonumber \\
  & -6(1-p_A)(1-p_B)(V_\tau^iA_\tau^j+V_\tau^jA_\tau^i)
(1-x^2)(1-\frac
{x^2}{6}) & \nonumber \\
 & [\alpha_A (1+p_A)(1-p_B)+\alpha_B (1+p_B)(1-p_A)]
\biggr), &
\end{eqnarray}
\begin{eqnarray}
  & \langle O_2^2\rangle =
{{1}\over{360x^2\sigma  }}\sum_{i,j}K_{ij}s m_\tau^2 & \nonumber \\ 
  & \hspace*{-5mm} 
  \biggl[ \biggl( 3[(1-p_A)^2+(1-p_B)^2][V_\tau^i V_\tau^j
(4+7x^2+4x^4)+A_\tau^iA_\tau^j 2(1-x^2)(2+3x^2)]
 & \nonumber \\
&  -2\alpha_A \alpha_B (1-p_A) (1-p_B)
[V_\tau^iV_\tau^j(4+7x^2+4x^4)+A_\tau^iA_\tau^j
4(1-x^2)^2]\biggr)
  & \nonumber \\
  & +6
\biggl( 6(1-x^2)(p_A-p_B)^2[V_\tau^iV_\tau^j(1+\frac{x^2}{4})
+A_\tau^iA_\tau^j(1-x^2)] &  \nonumber \\
  & \hspace*{-5mm}-(V_\tau^iA_\tau^j+V_\tau^jA_\tau^i)(1-x^2)(4+x^2)(p_A-p_B)
[\alpha_A(1-p_A)-\alpha_B(1-p_B)]\biggr)\biggr]. &
\end{eqnarray}

The expected uncertainty (1 standard deviation) on the measurement of 
{$d_{\tau}$} can then be calculated from  
\begin{equation}
\delta\,{\rm Re(Im)}
d_\tau^i= \frac{1}{c_{AB}^{a,i}}\:\frac{e}{\sqrt{s}}\frac{1}{\sqrt{N_{AB}}}\:
S_{a}, \hspace{0.8cm}
\begin{array}{l}\\
a=1(\rm{Re}),\,2(\rm{Im}),\\
i=\gamma,Z. 
\end{array}
\end{equation}
The electric charge of the electron is denoted by {$e$}. 
The coefficients {$c_{AB}^{a,i}$} are discussed in the next section.
{$N_{AB}$} is the number of events in the channel {$A\overline{B}$}
and {$\overline{A}B$}, and is given by
\begin{equation}
N_{AB}=N_{\tau^+\tau^-}B(\tau^-\rightarrow
A\nu_\tau )B(\tau^+\rightarrow \overline{B}\overline{\nu}_\tau),
\end{equation}
where we compute $N_{\tau^+\tau^-}$ for the design luminosities and
from the cross-section at the given energy and the polarizations.

\section{Results for TESLA}
\label{sec:TESLARes}
Here we present the results for the vector correlations
and their standard deviation at TESLA energies,
luminosities and polarizations, using the expressions of the
previous section.

We begin with the expression for the cross-section which is of the form
\begin{equation}
\sigma_1 (1 - P_e P_{\overline{e}}) + \sigma_2 (P_e -P_{\overline{e}})
\end{equation}
The values of {$\sigma_1$} and {$\sigma_2$} for the energies of interest
are given in Table 1.  Furthermore, in Fig.\ 1 and Fig.\ 2 we present profiles 
of the cross-sections as a function for {$P_e$} where the profiles correspond
to constant values of the positron polarization {$P_{\overline{e}}=0,0.3$}
and $0.6$.  We will, for the rest of the discussion, consider these to be
the reference polarizations for the positron beam.
Notice that the cross-section is larger when the {$e^+$} and {$e^-$} 
polarizations are opposite in sign, and then, it increases with {$e^+$}
polarization. This results in better sensitivities for the corresponding
cases.

\begin{table}[tb]
\begin{center}
\begin{tabular}{||c|c|c||}\hline
$\sqrt{s}$ (GeV) & $\sigma_1$ (fb) & $\sigma_2$ (fb) \\\hline
500 & 447.70 & -31.37 \\
800 & 174.03 &-11.94 \\ \hline
\end{tabular}
\caption{Coefficients for the cross-section in fb for energies of interest}
\label{tableofcrosssections}
\end{center}
\end{table}

Analogously, 
the expressions for the quantities {$c_{AB}^{a,i}$} and 
{$\langle O_a^2 \rangle$} may be schematically expressed as:
\vspace{-2mm}
\begin{equation}
c_{AB}^{a,i}=f\frac{C_1^{a,i}
(1 - P_e P_{\overline{e}}) + C_2^{a,i}(P_e -P_{\overline{e}})}
{\sigma_1 (1 - P_e P_{\overline{e}}) + \sigma_2 (P_e -P_{\overline{e}})},
\hspace{6mm}
\begin{array}{l}\\
a=1,2 \\
i=\gamma, Z
\end{array}
\end{equation}
\vspace{-1mm}
and
\begin{equation}
\langle{O_a^2}\rangle=f \frac{D_1^a (1 - P_e P_{\overline{e}}) +
 D_2^a(P_e -P_{\overline{e}})}
{\sigma_1 (1 - P_e P_{\overline{e}}) + \sigma_2(P_e -P_{\overline{e}})},
\hspace{10mm} a=1,2.
\end{equation}
respectively for the operators {$O_1$} and $O_2$, where
$f=4 \pi \alpha^2 (\hbar c)^2/3=9.818 \cdot 10^7$ GeV$^{2} \cdot$fb.
We have taken $\alpha^{-1}=128.87$.  
The quantities {$C_k^{a,i}$} and {$D_k^{a}$} are listed in Tables 2-5 for 
the different energies and channels.  
Note that the overall dimensions for $c_{AB}^{1,i}$ and
$c_{AB}^{2,i}, i=\gamma,Z$ are GeV$^2$ and GeV respectively and
that of $\langle O_1^2 \rangle$ and $\langle O_2^2 \rangle$ are
GeV$^4$ and GeV$^2$ respectively, which for brevity,
will not be mentioned in the rest of the discussion.

\begin{table}
\begin{center}
\hspace*{-1mm}
\begin{tabular}{ ||c | r | r | r | r | r | r ||}\hline
$\!$AB &   \multicolumn{1}{c|}{$C_1^{1,\gamma}$}
       &   \multicolumn{1}{c|}{$C_2^{1,\gamma}$}
       &   \multicolumn{1}{c|}{$C_1^{1,Z}\!$}
       &   \multicolumn{1}{c|}{$\!C_2^{1,Z}\!$}
       &   \multicolumn{1}{c|}{$\!D_1^1$}
       &   \multicolumn{1}{c|}{$D_2^1$}            \\\hline
$\pi\pi$&  $ \!2.70 \!\cdot\! 10^{-5} $&$2.94 \!\cdot\! 10^{-4}\!$&$\!
 -1.79 \!\cdot\! 10^{-4} \!$&$\! -2.12 \!\cdot\! 10^{-6} \!$&$\!
3.27 \!\cdot\! 10^{-2} \!$&$\!7.31 \!\cdot\! 10^{-3}\!$ \\
$\pi\rho\!$&$\! 6.67 \!\cdot\! 10^{-6} \!$&$\! 2.30 \!\cdot\! 10^{-4} \!$&$\!
 -1.41 \!\cdot\! 10^{-4} \!$&$\!  7.22 \!\cdot\! 10^{-6} \!$&$\!
2.92 \!\cdot\! 10^{-2}\!$&$\!3.28 \!\cdot\! 10^{-3}\!$ \\
$\rho\rho\!$&$\!1.20 \!\cdot\! 10^{-6}\!$&$\!1.32 \!\cdot\! 10^{-4}\!$&$\!
-8.08 \!\cdot\! 10^{-5} \!$&$\!  5.73 \!\cdot\! 10^{-6} \!$&$\!
2.47 \!\cdot\! 10^{-2}\!$&$\!1.10 \!\cdot\! 10^{-3}\!$ \\ \hline
\end{tabular}
\caption{List of coefficients for the operator $O_1$ for $\sqrt{s}=500$ GeV.}
\end{center}
\end{table}

\begin{table}
\begin{center}
\hspace*{-2mm}
\begin{tabular}{ ||c | r | r | r | r | r | r ||}\hline
$\!$AB &  \multicolumn{1}{c|}{$\!C_1^{2,\gamma}\!$}
       &  \multicolumn{1}{c|}{$\!C_2^{2,\gamma}\!$}
       &  \multicolumn{1}{c|}{$\!C_1^{2,Z}\!$}
       &  \multicolumn{1}{c|}{$\!C_2^{2,Z}\!$}
       &  \multicolumn{1}{c|}{$\!D_1^2\!$}
       &  \multicolumn{1}{c|}{$\!D_2^2$}             \\\hline
$\pi\pi  \!$&$\!    -8.61 \!\cdot\! 10^{-7} \!$&$\! 6.89 \!\cdot\! 10^{-8} \!$&$\!
 -8.46 \!\cdot\! 10^{-8} \!$&$\! 5.33 \!\cdot\! 10^{-7} \!$&$\!
   1.25 \!\cdot\! 10^{-2} \!$&$\!   -8.77 \!\cdot\! 10^{-4}\!$ \\
$\pi\rho \!$&$\!   -5.92 \!\cdot\! 10^{-7}  \!$&$\! 4.74 \!\cdot\! 10^{-8}  \!$&$\!
 -5.83 \!\cdot\! 10^{-8} \!$&$\! 3.66 \!\cdot\! 10^{-7} \!$&$\!
 1.44 \!\cdot\! 10^{-2} \!$&$\!  -2.38 \!\cdot\! 10^{-3}\!$ \\
$\rho\rho \!$&$\!  -3.24 \!\cdot\! 10^{-7}  \!$&$\! 2.59 \!\cdot\! 10^{-8} \!$&$\!
 -3.18 \!\cdot\! 10^{-8} \!$&$\! 2.10 \!\cdot\! 10^{-7} \!$&$\!
  1.17 \!\cdot\! 10^{-2}  \!$&$\! -8.17 \!\cdot\! 10^{-4}\!$\\ \hline
\end{tabular}
\caption{List of coefficients for the operator $O_2$ for $\sqrt{s}=500$ GeV.}
\end{center}
\end{table}

In order to get a feeling for the dependence of the quantities of interest
on the polarization, we illustrate {$\pi\pi$} channel.  
In Fig.\ 3, we present {$c^{1,\gamma}_{\pi\pi}$} as a function of $P_e$.
The sign of {$P_e$} is opposite to {$P_{\overline{e}}$} in order to
maximize the effects of longitudinal polarization.  
In Fig.\ 4, we have an analogous illustration
of $c^{1,Z}_{\pi\pi}$.  One may note that the curvature of the profiles
in Figs.\ 3 and 4 differ, which is dictated by the relative sizes and signs
of the coefficients {$C^{1,i}_{k}$} that enter
the final expressions for $c^{1,i}_{\pi\pi}$.  
In Fig.\ 5, we present profiles of the quantity {$S_1$} for the $\pi\pi$ 
channel.  In Fig.\ 6 and 7 we illustrate the
behavior of {$c^{2,\gamma}_{\pi\pi}$} and $c^{2,Z}_{\pi\pi}$.
For maximum electron polarization {$P_e = \pm 1$} all quantities become 
independent of {$P_{\overline{e}}$}, since the effective polarization 
parameter no longer depends on {$P_{\overline{e}}$} ($P = \pm 1$).
We do not illustrate the behavior of $S_2$ since this quantity 
is practically constant with a value of 52.40 in the entire range of the 
positron polarization.    
We have not plotted
the corresponding curves for the channels involving the $\rho$ and these
may be simply generated from the entries given in the tables.

We now discuss the limits achievable at TESLA with the design
luminosities and polarizations (we give 1 standard deviation uncertainties).
With a fixed value of electron and positron polarizations, one
can only obtain limits on linear combinations of the EDM and WDM.
Such limits would be defined by straight lines given by the equation
\begin{equation}
  \frac{\delta\,{\rm Re}d_\tau^\gamma}{a} +
  \frac{\delta\,{\rm Re}d_\tau^Z}{b}      = \pm 1
\end{equation}
\noindent for the limits arising from $O_1$ and by
\begin{equation}
  \frac{\delta\,{\rm Im}d_\tau^\gamma}{c} +
  \frac{\delta\,{\rm Im}d_\tau^Z}{d}      = \pm 1
\end{equation}
\noindent for the limits arising from $O_2$
where the numbers {$a$}, {$b$}, {$c$}, and {$d$} can be explicitly
computed for given polarizations and luminosities.
The value of {$a$} ($c$) is the sensitivity
to the  real (imaginary) part of the EDM
when the real (imaginary) part of the WDM is set to zero and
{$b$} ($d$) is the sensitivity to the WDM 
when the EDM is set to zero.

\begin{table}
\begin{center}
\begin{tabular}{ ||c | r | r | r | r | r | r ||}\hline
AB &   \multicolumn{1}{c|}{$C_1^{1,\gamma}$}       
   &   \multicolumn{1}{c|}{$C_2^{1,\gamma}$}
   &   \multicolumn{1}{c|}{$C_1^{1,Z}\!$}
   &   \multicolumn{1}{c|}{$\!C_2^{1,Z}\!$}
   &   \multicolumn{1}{c|}{$\!D_1^1   \!$}
   &   \multicolumn{1}{c|}{$\!    D_2^1$}        \\\hline
$\pi\pi \!$&$\!     1.65 \!\cdot\! 10^{-5} \!$&$\!  1.84 \!\cdot\! 10^{-4} \!$&$\!
 -1.10 \!\cdot\! 10^{-4} \!$&$\! -1.09 \!\cdot\! 10^{-6} \!$&$\!
  3.26 \!\cdot\! 10^{-2} \!$&$\!  7.17 \!\cdot\! 10^{-3} $ \\
$\pi\rho \!$&$\!    4.08 \!\cdot\! 10^{-6} \!$&$\!  1.44 \!\cdot\! 10^{-4} \!$&$\!
 -8.64 \!\cdot\! 10^{-5} \!$&$\! 4.47\!\cdot\! 10^{-6} \!$&$\!
  2.92 \!\cdot\! 10^{-2} \!$&$\!  3.22 \!\cdot\! 10^{-3} $ \\
$\rho\rho \!$&$\!   7.35 \!\cdot\! 10^{-7} \!$&$\!  8.22 \!\cdot\! 10^{-5} \!$&$\!
 -4.95 \!\cdot\! 10^{-4} \!$&$\! 3.52 \!\cdot\! 10^{-6} \!$&$\!
  2.46 \!\cdot\! 10^{-2} \!$&$\!  1.10 \!\cdot\! 10^{-3} $ \\ \hline
\end{tabular}
\caption{List of coefficients for the operator $O_1$ for $\sqrt{s}=800$ GeV.}
\end{center}
\end{table}

\begin{table}
\begin{center}
\hspace*{-2mm}
\begin{tabular}{ ||c | r | r | r | r | r | r ||}\hline
$\!$AB&   \multicolumn{1}{c|}{$C_1^{2,\gamma}$}
      &   \multicolumn{1}{c|}{$C_2^{2,\gamma}$}
      &   \multicolumn{1}{c|}{$C_1^{2,Z}\!$}
      &   \multicolumn{1}{c|}{$\!C_2^{2,Z}\!$}
      &   \multicolumn{1}{c|}{$\!D_1^2$}
      &   \multicolumn{1}{c|}{$D_2^2$}        \\\hline
$\pi\pi \!$&$\!     -3.29 \!\cdot\! 10^{-7} \!$&$\!   2.63 \!\cdot\! 10^{-8} \!$&$\!
 -3.17 \!\cdot\! 10^{-8} \!$&$\! 2.00 \!\cdot\! 10^{-7} \!$&$\!
  1.25 \!\cdot\! 10^{-2} \!$&$\!  -8.54 \!\cdot\! 10^{-4} \!$ \\
$\pi\rho \!$&$\!   -2.26 \!\cdot\! 10^{-7} \!$&$\!  1.81 \!\cdot\! 10^{-8} \!$&$\!
 -2.18 \!\cdot\! 10^{-8} \!$&$\! 1.37 \!\cdot\! 10^{-7} \!$&$\!
  1.43 \!\cdot\! 10^{-2} \!$&$\!  -2.32 \!\cdot\! 10^{-3}\!$ \\
$\rho\rho \!$&$\!   -1.24 \!\cdot\! 10^{-7} \!$&$\!  9.92 \!\cdot\! 10^{-9} \!$&$\!
 -1.19 \!\cdot\! 10^{-8} \!$&$\! 7.51 \!\cdot\! 10^{-8} \!$&$\!
  1.16 \!\cdot\! 10^{-2} \!$&$\!   -7.96 \!\cdot\! 10^{-4} \!$\\ \hline
\end{tabular}
\caption{List of coefficients for the operator $O_2$ for $\sqrt{s}=800$ GeV.}
\end{center}
\end{table}

The quantities {$a$}, {$b$}, {$c$}, and {$d$} are plotted in Figs.\ 
8-11 as functions of {$P_e$} in the vicinity of {$P_e= -0.8$}, 
for the reference values of {$P_{\overline{e}}$}. 
In all cases, the variation with $P_e$ and
{$P_{\overline{e}}$} is slow in the range we have chosen, since we are
already close to the maximum effective polarization.   
Many models predict a significantly larger EDM than WDM. 
In this case the limits achievable are precisely {$a$} and {$c$}
on the real and imaginary parts of the EDM.
If this is not the case, the EDM and WDM can be disentangled by switching
the relative signs of {$P_e$} and {$P_{\overline{e}}$}.
The limits achievable for equal luminosities for both settings 
are given in Tab.\ 6.
The pairs {$(\cal{A},\cal{B})$} and {$(-\cal{A},-\cal{B})$} give the vertices
in the Re $d^\gamma_\tau$-Re $d^Z_\tau$ plane and {$(\cal{C},\cal{D})$} and 
{$(-\cal{C},-\cal{D})$}
in the analogous Im $d^\gamma_\tau$-Im $d^Z_\tau$ plane.

We also present below a table of numbers computed for the contribution
from the helicity flip bremsstrahlung to the operator $O_2$.  We
schematically express it as
\begin{equation}
\langle O_2 \rangle_{SM}= {- A'(P_e+P_{\overline{e}})\over \sigma_1
(1-P_e P_{\overline{e}})+ \sigma_2(P_e-P_{\overline{e}})}
\end{equation}
where the quantity $\langle O_2 \rangle_{SM}$ has 
the overall dimension of GeV, and the corresponding coefficient $A'$ is
tabulated in Tab.\ 7. 
The cross-sections {$\sigma_i$} are given in Tab.\ 1.
Even with extreme values of {$P_e$} and {$P_{\overline{e}}$}, 
we get a number of order 1, to be contrasted with $S_2 \simeq
50$ for the $\pi\pi$ channel thereby rendering this background negligible.
\pagebreak

\begin{table}
  \begin{center}
    \begin{tabular}{||c||r|r||r|r||}\hline
    $\sqrt{s}$ GeV & \multicolumn{1}{c|}{$\cal{A}$} 
    		   & \multicolumn{1}{c|}{$\cal{B}$}
    		   & \multicolumn{1}{c|}{$\cal{C}$}
    		   & \multicolumn{1}{c|}{$\cal{D}$} \\ \hline
    500 & $ 3.73 \cdot 10^{-20}$ & $-5.90 \cdot 10^{-19}$ 
    	& $-9.31 \cdot 10^{-17}$ & $ 3.15 \cdot 10^{-18}$ \\
    	& $-3.84 \cdot 10^{-19}$ & $ 1.03 \cdot 10^{-20}$ 
    	& $ 1.05 \cdot 10^{-17}$ & $-1.60 \cdot 10^{-16}$ \\ \hline
    800 & $ 2.63 \cdot 10^{-20}$ & $-4.25 \cdot 10^{-19}$ 
    	& $-1.07 \cdot 10^{-16}$ & $ 3.90 \cdot 10^{-18}$ \\
    	& $-2.71 \cdot 10^{-19}$ & $ 7.39 \cdot 10^{-21}$ 
    	& $ 1.22 \cdot 10^{-17}$ & $-1.88 \cdot 10^{-16}$ \\ \hline
    \end{tabular}
    \caption{Sensitivities achievable when signs of electron and
  	     positron polarizations are interchanged.}
    \label{tab:ind}
  \end{center}
\end{table}
  
\begin{table}
\begin{center}
\begin{tabular}{||c | c | c||} \hline
AB         & $\sqrt{s}$=500 GeV & $\sqrt{s}$=800 GeV  \\ \hline
$\pi\pi$   & 338                & 243   \\
$\pi\rho$  & 343                & 242   \\
$\rho\rho$ & 349                & 241   \\ \hline
\end{tabular}
\caption{Coefficient $A'$ of helicity-flip bremsstrahlung to $\langle O_2
\rangle_{SM}$.}
\end{center}
\end{table} 

TESLA can also be operated at the Z-pole.
It is expected that sufficient integrated luminosity will
be available to generate as many as $10^9$ Z bosons.
A simple scaling of the limits obtained in ref.\ \cite{basdr1} with the 
effective
polarization parameter {$P\simeq 0.95$} could yield 1 s.d. limits of
$3\times 10^{-19}$~{\ecm} on the real part, and $10^{-18}$~{\ecm} on the 
imaginary part of the weak dipole moment.

\section{Experimental Aspects}
\label{sec:ExpAspects}

Up to this point we have done analytic calculations of vector correlations
for two decay channels of the {$\tau$} lepton. 
This gives us a realistic estimate of the limits that can be achieved on the
EDM and WDM with these decays with a perfect detector. 
Here we want to discuss possibilities of improving the sensitivity further and 
we try to estimate by how much the result will be weakened due to detector
effects in a real experiment. 
We will try to extrapolate the size of the detector effects from the 
experience from the LEP experiments and SLD \cite{dtauLEP}.

\begin{itemize}
  \item In the calculations we have taken into account the decays
        into {$\pi$} and {$\rho$}. These are the two most important channels. 
        Together they make up {$13\;\%$} of the total branching
        ratio of {$\tau$} pairs. More channels can be added. 
        The decay modes {\tauppp} and {\taupppn} and the two leptonic 
        decays {\taue} and {\taumu} have been used at LEP/SLC. 
        This increases the potential fraction of the sample used in the 
        analysis to {$82\;\%$}.
        However, these additional channels have a lower sensitivity 
        to CP violation. 
        Taking this into account the theoretically achievable sensitivity
        increased by a factor of 2.8 at the Z-pole. There will probably
        be a similar factor at TESLA.
  \item The vector correlations discussed here receive contributions from 
        the most important part of the cross-section. They are more 
        sensitive than the tensor correlations initially discussed in 
        \cite{Nachtmann}, if beam polarization is available.
        With so-called optimal observables \cite{NachtmannOptimal} the 
        sensitivity from every part of the cross-section can be exploited.  
        The experiments at LEP/SLC started off using the tensor
        correlations. When they later moved to the optimal observables,
        the sensitivity improved by roughly an order of magnitude. 
        Also here we expect a gain in sensitivity from the optimal 
        observables, although the situations are not directly comparable.
  \item The real detector will have a finite energy and momentum resolution
        for the decay products of the {$\tau$} leptons. This affects the
        calculation of the observables and reduces the sensitivity.
        The reduction of sensitivity at LEP/SLC was less than {$10\;\%$} 
        for events without {\pio} in the final state and in the order of
        {$10\;\%$} for events with {\pio} mesons. Despite the higher 
        energies, the relative momentum resolution of the TESLA detector 
        for charged particles should be slightly better than that at 
        LEP/SLC.
        The energy resolution of the $\pio$s (causing the main loss of 
        sensitivity at LEP/SLC) should be substantially better at TESLA. 
        We assume that the loss of sensitivity due to finite energy
        and momentum resolution will not be larger than {$10\;\%$}.
  \item In a real experiment the event samples selected will contain 
        background from misidentified {$\tau$} decays and also from non
        {$\tau$} pair events. The background dilutes the signal and 
        reduces the sensitivity. This reduction of sensitivity was 
        between 1 and 10 percent at LEP/SLC for the different decay
        channels. At higher center-of-mass energies at TESLA the higher
        Lorentz boost of the {$\tau$} leptons makes the identification of
        their decay channels more difficult. But then we also expect the 
        TESLA detector to have a better performance. Especially the 
        high granularity of the proposed silicon-tungsten electromagnetic
        calorimeter will simplify {$\tau$} analysis. 
        Overall we don't expect to lose more than {$10\;\%$} in sensitivity
        due to background.
  \item The real detector will not have {$100\;\%$} efficiency in the 
        identification of the signal. At LEP/SLC typical values for the 
        overall efficiencies ranged between 50 and {$90\;\%$} for the 
        various decay modes of the {$\tau$} pairs. 
        The TESLA detector should perform at least as good as the 
        LEP/SLC detectors. 
\end{itemize}

Without a detailed study based on a full analysis of events with full detector
simulation it is impossible to tell by how much the limit achievable in the
real experiment will differ from our analytic estimate. From the argument 
above we conclude that the achievable limit will not be worse than our analytic
estimate. It will probably be better by a factor of a few.

\section{Discussion and Conclusions}
\label{sec:Conclusions}

In the present work, we have considered vector correlations among the
momenta of charged decay products of the {$\tau^\pm$} in two-body decays
into final states with {$\pi$} and {$\rho$}.
By considering CP-odd operators which are T-odd ($O_1$) and T-even ($O_2$) 
it is possible to probe the real and imaginary parts of CP violating dipole 
form factors of the {$\tau$} lepton.
We expect the following precision (1 standard deviations) with the expected
beam polarizations of {$-80\ \%$} for the electron beam and {$60\ \%$} for 
the positron beam (in units of {\ecm}).
\begin{center}
  \begin{tabular}{ccrrrr}
    $\sqrt{s}$&$\int \mathrm{dt} \cal{L} $
           & ${\rm Re}\;d^\gamma_\tau$~~~ & ${\rm Im}\;d^\gamma_\tau$~~~ & 
             ${\rm Re}\;d^Z_\tau$~~~      & ${\rm Im}\;d^Z_\tau$~~~      \\
    500 GeV&{$~500\ \mathrm{fb}^{-1}$}& $3.8 \cdot 10^{-19}$&
                                        $0.9 \cdot 10^{-16}$&
                                        $5.4 \cdot 10^{-19}$&
                                        $1.4 \cdot 10^{-16}$\\
    800 GeV&{$1000\ \mathrm{fb}^{-1}$}& $2.7 \cdot 10^{-19}$&
                                        $1.1 \cdot 10^{-16}$&
                                        $3.9 \cdot 10^{-19}$&
                                        $1.7 \cdot 10^{-16}$\\
  \end{tabular}
\end{center}
In these determinations the limits on the electric dipole moment are obtained 
assuming the weak dipole moment is zero and vice-versa.
It is possible to disentangle the individual limits by switching the beam
polarizations (see Tab.\ \ref{tab:ind}). 

A priori it cannot be said
by how much better tensor correlations of the type considered in \cite{bno}  
will fare when polarization is included and conclusions cannot be drawn
unless they are fully studied.  However, it must be noted that the
sensitivities we have reported here are significantly superior to those
obtainable from the tensor correlations with no
beam polarization.

The vector correlations are significantly enhanced due to longitudinal 
polarization of the beams. The precision achievable without polarization
would be at least an order of magnitude reduced. 
Since the effective polarization parameter is already close to unity with the 
designed electron polarization alone, the gain from positron polarization even 
at {$60\:\%$} improves the sensitivity by a factor of 2 at most.
The measurements are not very sensitive to the precise value of the 
polarizations. A control on the polarization at the level of a few percent
is sufficient. 

\section{Acknowledgments}
It is our pleasure to thank Prof. O. Nachtmann for valuable discussions.


\begin{figure}[p!]
  \begin{center}
    \epsfig{figure=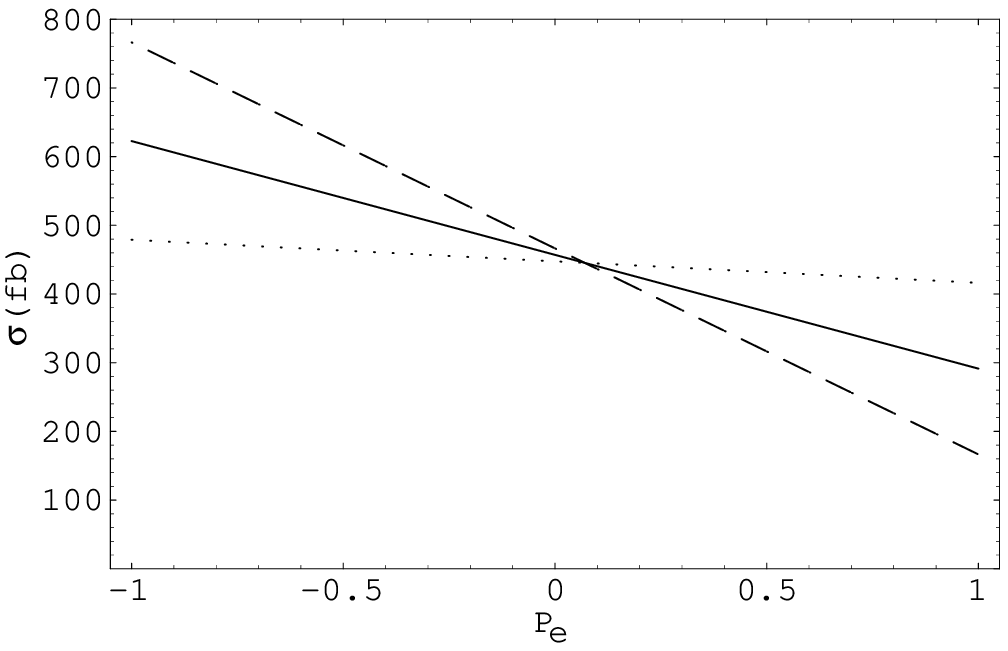,height=16cm, width=12cm}
    \caption{Values of the cross-section
             for $P_{\overline{e}}=0,0.3,0.6$ (dotted, solid
             and dashed) as a function of $P_e$, with $\sqrt{s}=500$ GeV.}
  \end{center}
\end{figure}
\begin{figure}[p!]
  \begin{center}
    \epsfig{figure=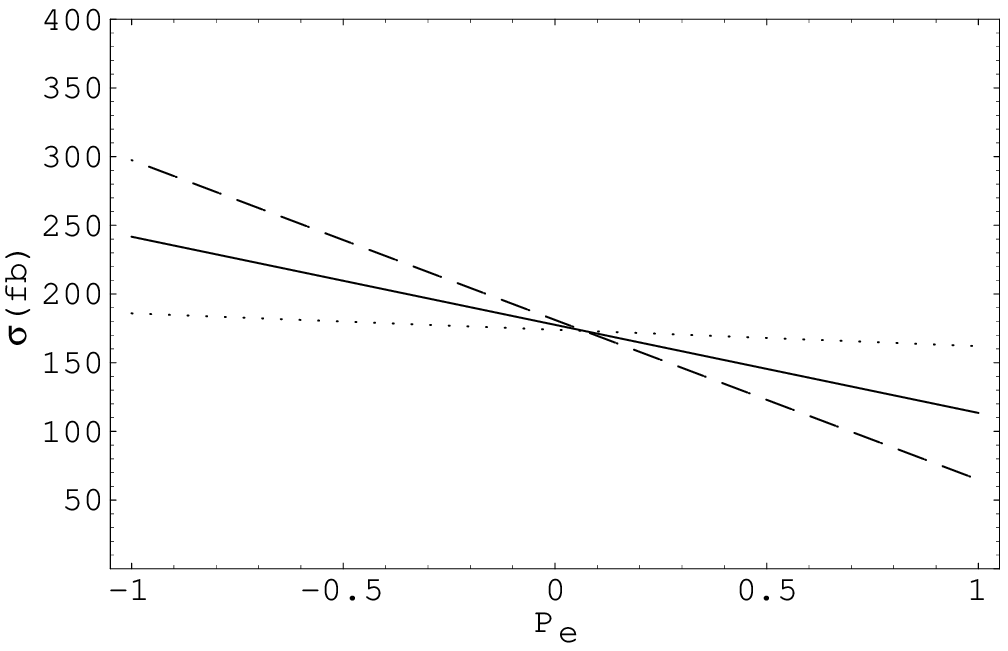,height=16cm, width=12cm}
    \caption{Values of the cross-section
             for $P_{\overline{e}}=0,0.3,0.6$ (dotted, solid
             and dashed) as a function of $P_e$, with $\sqrt{s}=800$ GeV.}
  \end{center}
\end{figure}
\begin{figure}[p!]
  \begin{center}
    \epsfig{figure=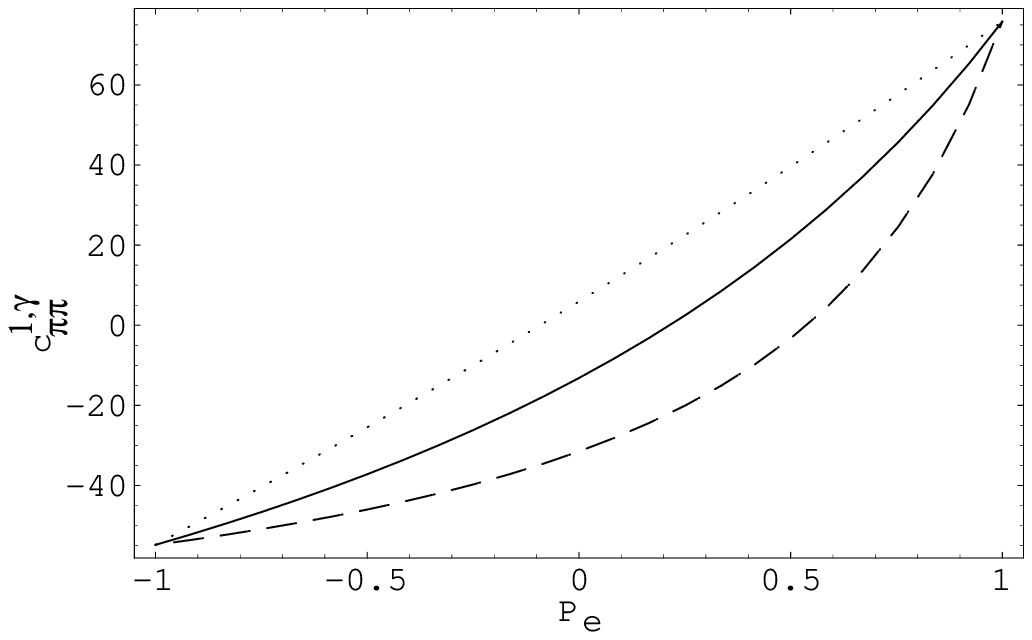,height=16cm, width=12cm}
    \caption{Values of $c_{\pi\pi}^{1,\gamma}$
             for $P_{\overline{e}}=0,0.3,0.6$ (dotted, solid
             and dashed) as a function of $P_e$, with $\sqrt{s}=500$ GeV.}
  \end{center}
\end{figure}
\begin{figure}[p!]
  \begin{center}
    \epsfig{figure=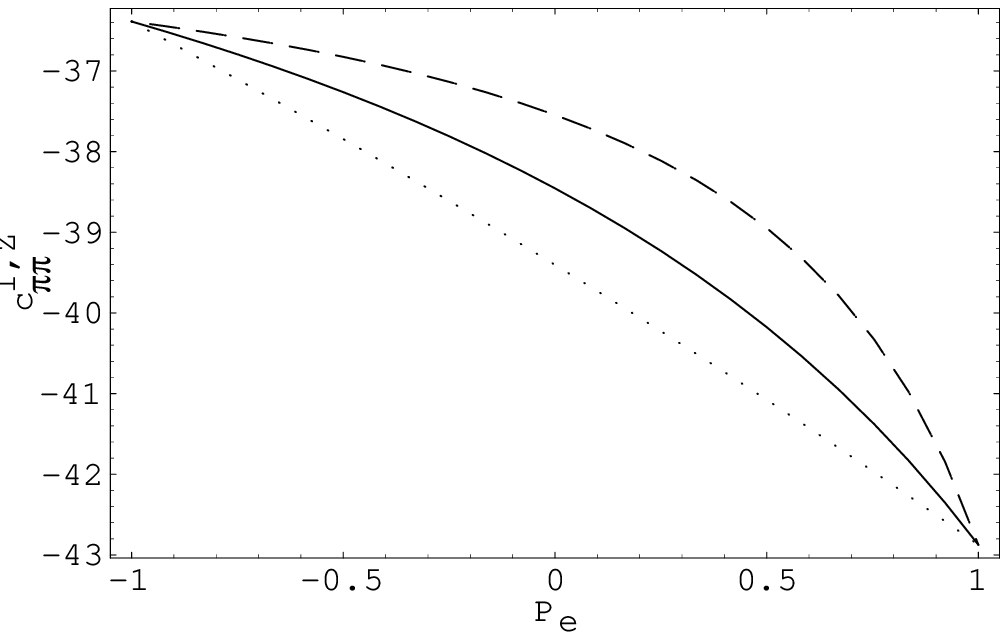,height=16cm, width=12cm}
    \caption{Values of $c_{\pi\pi}^{1,Z}$
             for $P_{\overline{e}}=0,0.3,0.6$ (dotted, solid
             and dashed) as a function of $P_e$, with $\sqrt{s}=500$ GeV.}
  \end{center}
\end{figure}
\begin{figure}[p!]
  \begin{center}
    \epsfig{figure=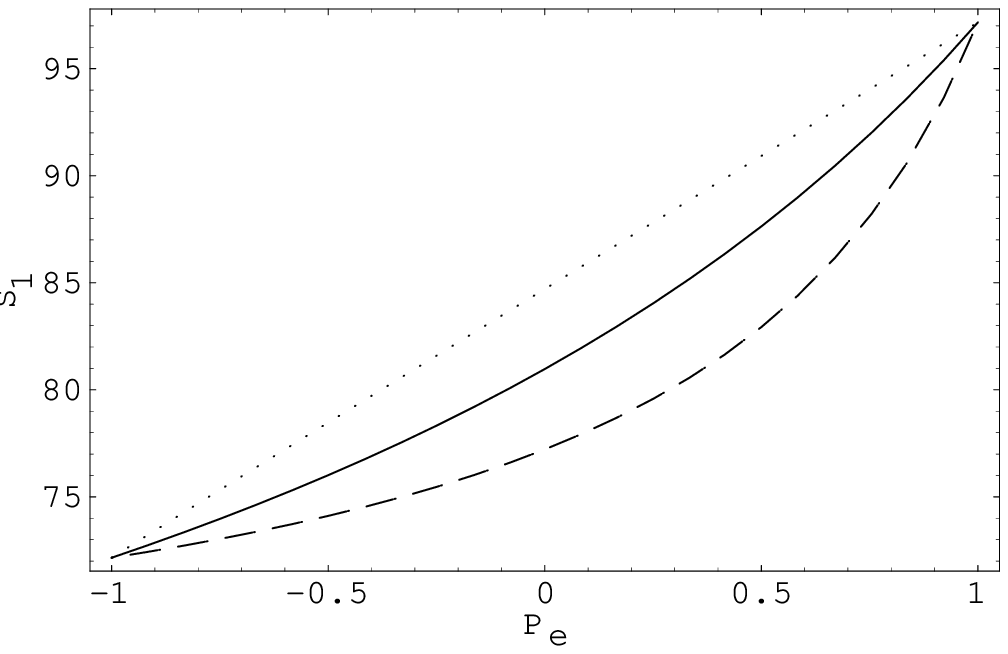,height=16cm, width=12cm}
    \caption{Values of $S_{1}$ for $P_{\overline{e}}=0,0.3,0.6$ (dotted, solid
             and dashed) as a function of $P_e$, with $\sqrt{s}=500$ GeV.}
  \end{center}
\end{figure}
\begin{figure}[p!]
  \begin{center}
    \epsfig{figure=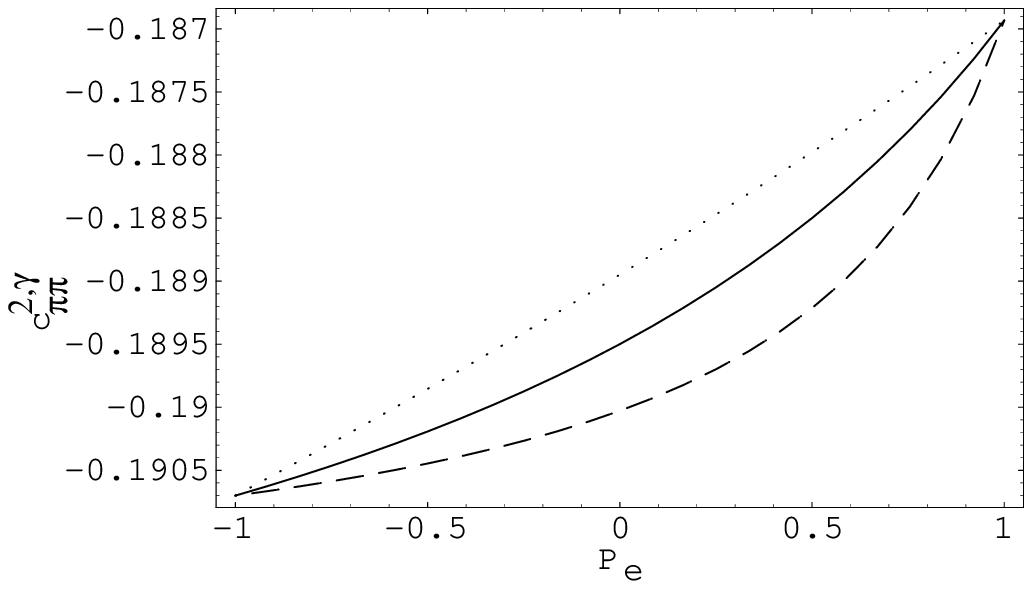,height=16cm, width=12cm}
    \caption{Values of $c_{\pi\pi}^{2,\gamma}$
             for $P_{\overline{e}}=0,0.3,0.6$ (dotted, solid
             and dashed) as a function of $P_e$, with $\sqrt{s}=500$ GeV.}
  \end{center}
\end{figure}
\begin{figure}[p!]
  \begin{center}
    \epsfig{figure=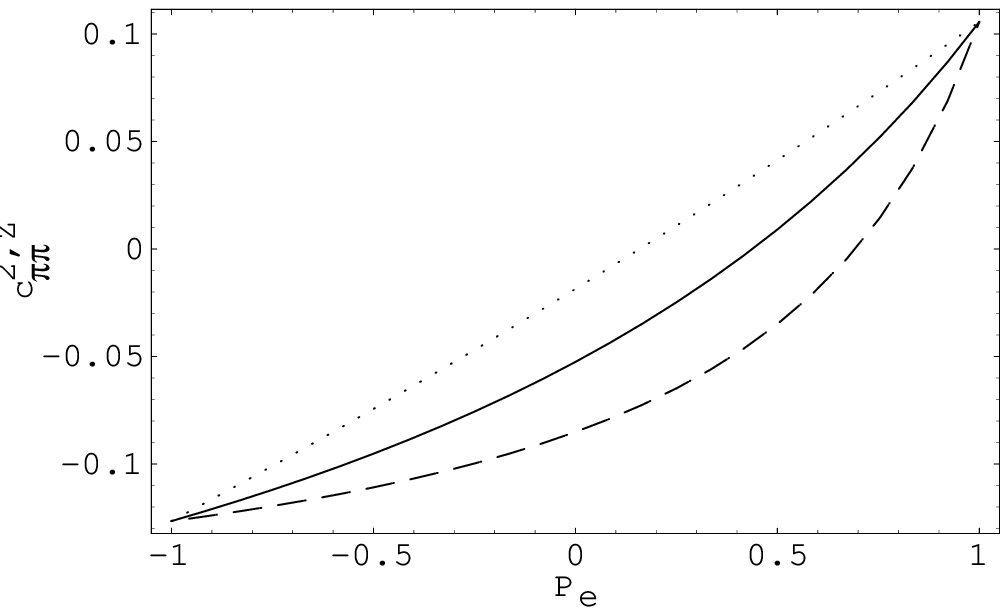,height=16cm, width=12cm}
    \caption{Values of $c_{\pi\pi}^{2, Z}$
             for $P_{\overline{e}}=0,0.3,0.6$ (dotted, solid
             and dashed) as a function of $P_e$, with $\sqrt{s}=500$ GeV.}
  \end{center}
\end{figure}
%
%
\begin{figure}[p!]
  \begin{center}
    \epsfig{figure=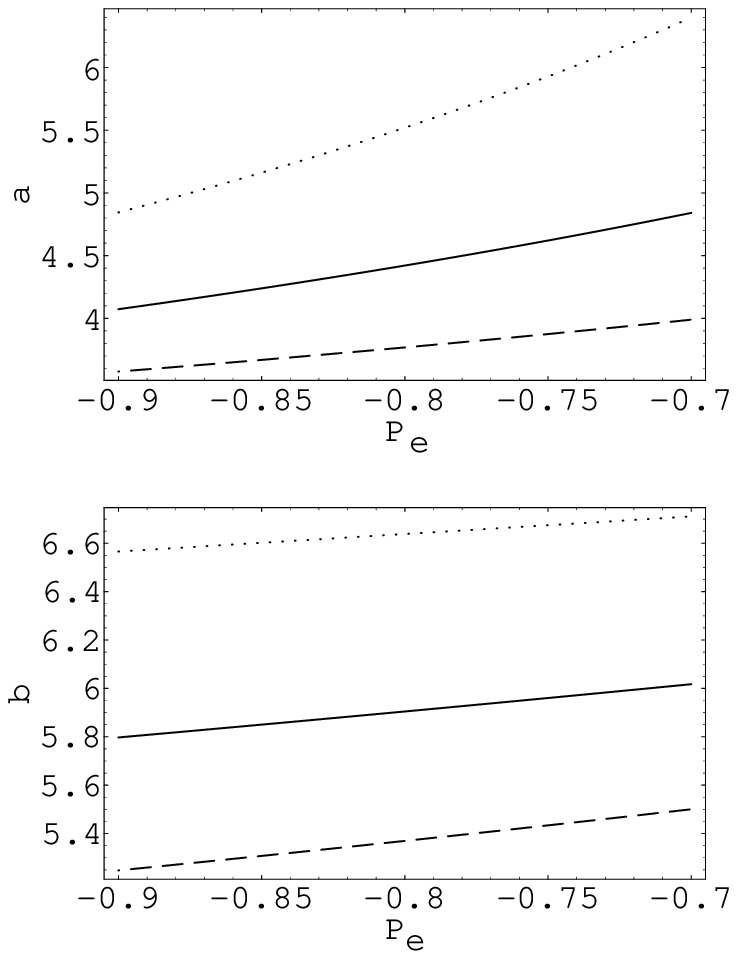,height=16cm, width=12cm}
    \caption{Values of $a$ and $b$ (in units of $10^{-19}\;\mbox{\ecm}$) for
             $P_{\overline{e}}=0,0.3,0.6$ (dotted, solid and dashed) as a
             function of {$P_e$} in the vicinity of the expected value of 
             -0.8 for $\int dt\cdot{\cal{L}}= 500 \, {\rm fb}^{-1}$ and 
             $\sqrt{s}=500$ GeV.}
  \end{center}
\end{figure}
\begin{figure}[p!]
  \begin{center}
    \epsfig{figure=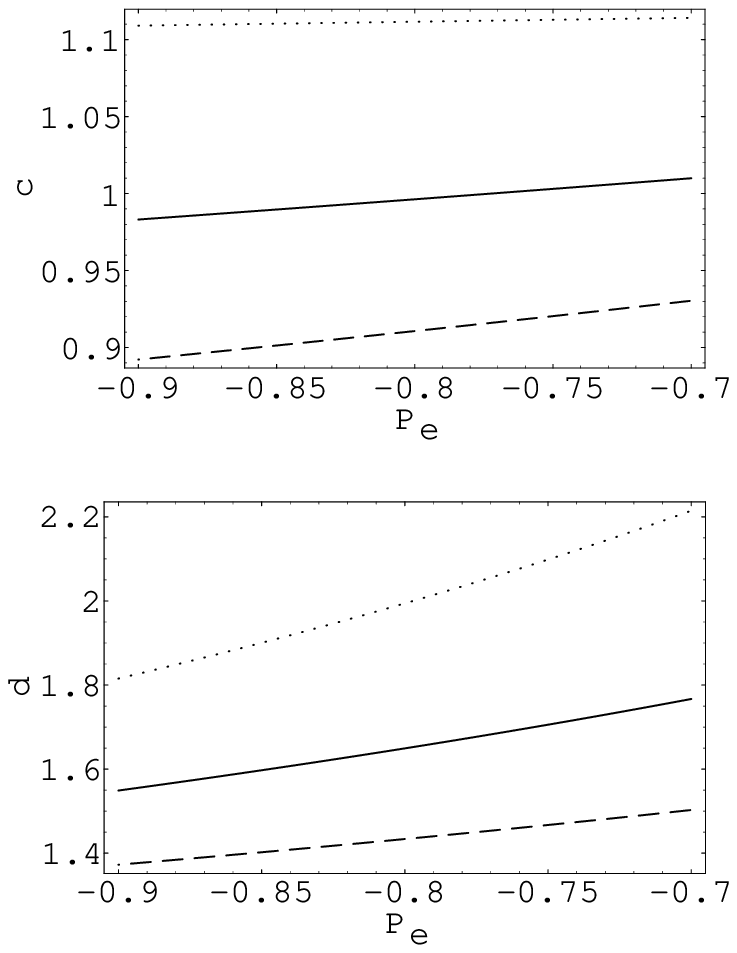,height=16cm, width=12cm}
    \caption{Values of $c$ and $d$ (in units of $10^{-16}\;\mbox{\ecm}$)
             for $P_{\overline{e}}=0,0.3,0.6$ (dotted, solid
             and dashed) as a
             function of {$P_e$} in the vicinity of the expected value of 
             -0.8 for $\int dt\cdot{\cal{L}}=
             500 \, {\rm fb}^{-1}$ and $\sqrt{s}=500$ GeV.}
  \end{center}
\end{figure}
\begin{figure}[p!]
  \begin{center}
    \epsfig{figure=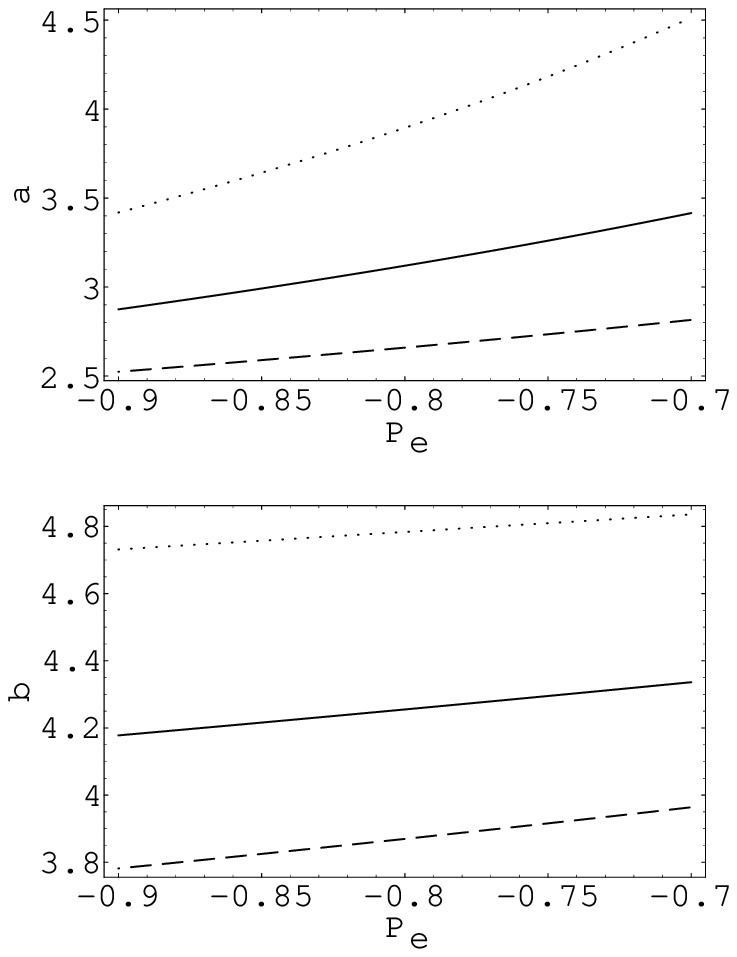,height=16cm, width=12cm}
    \caption{Values of $a$ and $b$ (in units of $10^{-19}\;\mbox{\ecm}$)
             for $P_{\overline{e}}=0,0.3,0.6$ (dotted, solid
             and dashed) as a
             function of {$P_e$} in the vicinity of the expected value of 
             -0.8 for $\int dt\cdot{\cal{L}}=
             1000 \, {\rm fb}^{-1}$ and $\sqrt{s}=800$ GeV.}
  \end{center}
\end{figure}
\begin{figure}[p!]
  \begin{center}
    \epsfig{figure=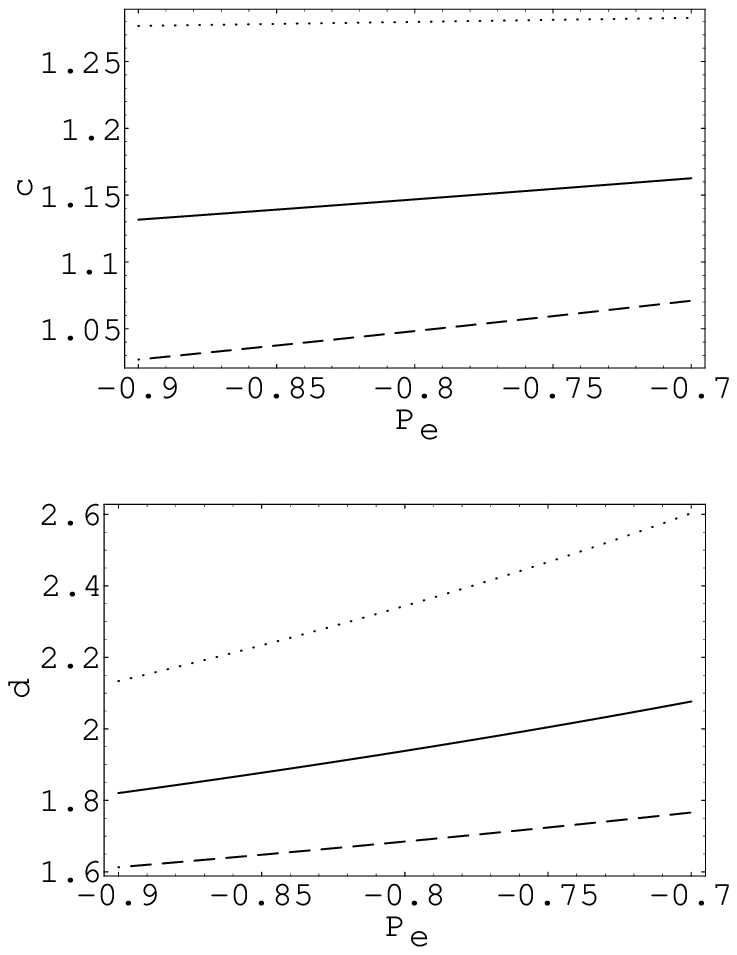,height=16cm, width=12cm}
    \caption{Values of $c$ and $d$ (in units of $10^{-16}\;\mbox{\ecm}$)
             for $P_{\overline{e}}=0,0.3,0.6$ (dotted, solid
             and dashed) as a
             function of {$P_e$} in the vicinity of the expected value of 
             -0.8 for $\int dt\cdot{\cal{L}}=
             1000 \, {\rm fb}^{-1}$ and $\sqrt{s}=800$ GeV.}
  \end{center}
\end{figure}

\end{document}